\RequirePackage{lineno}
\documentclass[aps,twocolumn,showpacs,byrevtex,prl,reprint]{revtex4-1}

\usepackage{graphicx}% Include figure files
\usepackage{dcolumn}% Align table columns on decimal point
\usepackage{bm}% bold math
\usepackage{rotating}
\usepackage{epstopdf}
\usepackage{color}
\usepackage{verbatim} %for comment
\usepackage{multirow}
\usepackage[abs]{overpic}

\newcommand{\PreserveBackslash}[1]{\let\temp=\\#1\let\\=\temp}
\newcolumntype{C}[1]{>{\PreserveBackslash\centering}p{#1}}
\newcolumntype{R}[1]{>{\PreserveBackslash\raggedleft}p{#1}}
\newcolumntype{L}[1]{>{\PreserveBackslash\raggedright}p{#1}}

\newcommand{\EE}{e^+e^-}

\newcommand{\psip}{\psi(3686)}

\newcommand{\jpsi}{J/\psi}
\newcommand{\too}{\rightarrow}

\begin{document}
%\linenumbers
\graphicspath{{figure/}}
\DeclareGraphicsExtensions{.eps,.png,.ps}

\title{\quad\\[0.0cm] \boldmath Observation of $\psip \too e^{+}e^{-}\chi_{cJ}$ and $\chi_{cJ} \too e^{+}e^{-}J/\psi$}

\author{
\begin{small}
    \begin{center}
      M.~Ablikim$^{1}$, M.~N.~Achasov$^{9,e}$, X.~C.~Ai$^{1}$,
      O.~Albayrak$^{5}$, M.~Albrecht$^{4}$, D.~J.~Ambrose$^{44}$,
      A.~Amoroso$^{49A,49C}$, F.~F.~An$^{1}$, Q.~An$^{46,a}$,
      J.~Z.~Bai$^{1}$
% ...
      \\
      \vspace{0.2cm}
      (BESIII Collaboration)\\
      \vspace{0.2cm} {\it
        $^{1}$ Institute of High Energy Physics, Beijing 100049, People's Republic of China\\
% ...
$^{4}$ Bochum Ruhr-University, D-44780 Bochum, Germany\\
        $^{5}$ Carnegie Mellon University, Pittsburgh, Pennsylvania 15213, USA\\
% ...
$^{9}$ G.I. Budker Institute of Nuclear Physics SB RAS (BINP), Novosibirsk 630090, Russia\\
% ...
$^{44}$ University of Rochester, Rochester, New York 14627, USA\\
% ...
$^{46}$ University of Science and Technology of China, Hefei 230026, People's Republic of China\\
% ...
$^{49}$ (A)University of Turin, I-10125, Turin, Italy; (B)University of Eastern Piedmont, I-15121, Alessandria, Italy; (C)INFN, I-10125, Turin, Italy\\
% ...
$^{a}$ Also at State Key Laboratory of Particle Detection and Electronics, Beijing 100049, Hefei 230026, People's Republic of China\\
% ...
$^{e}$ Also at the Novosibirsk State University, Novosibirsk, 630090, Russia\\
% ...
      }\end{center}
    \vspace{0.4cm}
    \end{small}
  }

%%% Local Variables:
%%% mode: latex
%%% TeX-master: "hc"
%%% End:

\begin{abstract}
 Using $4.479 \times 10^{8}$ $\psip$ events collected with the BESIII detector, we search for the decays $\psip \too e^{+}e^{-}\chi_{cJ}$ and $\chi_{cJ} \too e^{+}e^{-}J/\psi$, where $J=0,1,2$.
 The decays $\psip \too e^{+}e^{-}\chi_{cJ}$ and $\chi_{cJ} \too e^{+}e^{-}J/\psi$ are observed for the first time. The measured branching fractions are $\mathcal{B}(\psip \too e^{+}e^{-}\chi_{cJ}) = (11.7 \pm 2.5 \pm 1.0)\times10^{-4}$, $(8.6 \pm 0.3 \pm 0.6)\times10^{-4}$, $(6.9 \pm 0.5 \pm 0.6)\times10^{-4}$ for $J=0, 1, 2$, and $\mathcal{B}(\chi_{cJ} \too e^{+}e^{-}J/\psi) = (1.51 \pm 0.30 \pm 0.13)\times10^{-4}$, $(3.73 \pm 0.09 \pm 0.25)\times10^{-3}$, $(2.48 \pm 0.08 \pm 0.16)\times10^{-3}$ for $J=0, 1, 2$, respectively.
 The ratios of the branching fractions $\frac{\mathcal{B}(\psi(3686) \too e^{+}e^{-}\chi_{cJ})}{\mathcal{B}(\psi(3686) \too \gamma\chi_{cJ})}$ and $\frac{\mathcal{B}(\chi_{cJ} \too e^{+}e^{-}J/\psi)}{\mathcal{B}(\chi_{cJ} \too \gamma J/\psi)}$ are also reported. Also, the $\alpha$ values of helicity
angular distributions of the $\EE$ pair are determined for $\psip \too e^{+}e^{-}\chi_{c1,2}$ and $\chi_{c1,2} \too e^{+}e^{-}J/\psi$.

\end{abstract}

\pacs{13.20.Gd, 13.40.Hq, 14.40.Pq}

\maketitle
%%%%%%%%%%%%%%%%%%%%%%%%%%%%%%%%%%%%%%%%%%%%%%%%%%%%%%%%%%%%%%%%%%%%%%%%%%%%%%
%%%%%%%%%%%%%%%%%%%%%%%%%%%%%%%%%%%%%%%%%%%%%%%%%%%%%%%%%%%%%%%%%%%%%%%%%%%%%%
Study of electromagnetic (EM) Dalitz decays~\cite{dalitz}, in which a virtual photon is internally converted into an $e^{+}e^{-}$ pair, plays an important role in revealing the structure of hadrons and the interactions between photons and hadrons~\cite{theory}.
Such decays are widely observed in the light-quark meson sector, for example, $\eta' \too \gamma e^{+}e^{-}, \eta' \too \omega e^{+}e^{-},$ and $\phi \too \eta e^{+}e^{-}$~\cite{pdg}. However, the analogous transitions in charmonium decays have not yet been studied.
Although the potential quark model has successfully described the low-lying charmonium states with high precisions, there are still puzzling discrepancies in the decay branching fractions $\mathcal{B}(\psip \too \gamma\chi_{cJ})$ between the experimental results~\cite{pdg} where the higher-order multipole amplitudes are ignored and the various theoretical predictions~\cite{model1,model2,model3,model4}.
Throughout this Letter, $\chi_{cJ}$ refers to $\chi_{c0,1,2}$.
While recently the BESIII experiment confirms that the contributions from the higher-order multipole amplitudes in $\psip \too \gamma\chi_{cJ}$ are small~\cite{gaoq}, the E1 contribution is dominant.
Therefore, it is of great interest to measure the EM transition $\psip \too e^{+}e^{-}\chi_{cJ}$ and $\chi_{cJ} \too e^{+}e^{-}J/\psi$.

The EM Dalitz decays in charmonium transitions, such as $\psip \too e^{+}e^{-}\chi_{cJ}$ or $\chi_{cJ} \too e^{+}e^{-}J/\psi$, have access to the EM transition form factors (TFFs) of these charmonium states. The $q^2$-dependence of charmonium TFFs can provide additional information on the interactions between the charmonium states and the electromagnetic field, where $q^{2}$ is the square of the invariant mass of the $e^+e^-$ pair, and serve as a sensitive probe to their internal structures.
Furthermore, the
$q^2$-dependent TFF can possibly distinguish the transition mechanisms based on the $c\bar{c}$ scenario and other solutions which alter the simple quark model picture.
We emphasize that the $q^2$-dependent TFF can also serve as an useful probe for exotic hadron structures based on different models. One example is that with the precise measurement of the radiative decay of $X(3872)\too\EE J/\psi$ and $X(3872)\to\EE\psip$ in the future, we can pin down the intrinsic structure of $X(3872)$ by comparing the experimental measurement of the $q^2$-dependence of TFF with different model calculations.
The nature of $X(3872)$, namely whether it is a compact charmonium, multiquark state with quark clustering, or hadronic molecule~\cite{tetraquark1,tetraquark2,molecular2,X3872a,X3872b}, can possibly be disentangled by the $q^2$-dependence of its TFF. 

In this Letter, we report the observation of the EM Dalitz decays $\psip \too e^{+}e^{-}\chi_{cJ}$ and $\chi_{cJ} \too e^{+}e^{-}J/\psi$ by analysing the cascade decays $\psip \too e^{+}e^{-}\chi_{cJ}, \chi_{cJ} \too \gamma J/\psi$ and $\psip \too \gamma\chi_{cJ}, \chi_{cJ} \too e^{+}e^{-} J/\psi$, respectively. Here, the $J/\psi$ is reconstructed in its decay to an $e^{+}e^{-}$ or $\mu^{+}\mu^{-}$ pair.
The two cascade decays studied have the same final state: four leptons and a single photon.
The analysis uses a data sample of $4.479 \times 10^{8}$ $\psip$ events~\cite{totalnumber,totalnumber2} taken at a center-of-mass energy $\sqrt{s} = 3.686$ GeV collected with the BESIII detector~\cite{besiii} operating at the BEPCII~\cite{bepcii} storage ring in 2009 and 2012.
In addition, a data sample corresponding to an integrated luminosity of 44 pb$^{-1}$, taken at a center-of-mass energy $\sqrt{s} = 3.65$ GeV~\cite{data3650}, is used to estimate the background from continuum processes.

The BESIII detector~\cite{besiii} has a geometrical acceptance of 93$\%$ of the total 4$\pi$ solid angle. A small-cell helium-based main drift chamber (MDC) provides momentum measurements of charged particles with resolution of 0.5$\%$ at 1 GeV/$c$. The MDC also supplies an energy loss ($dE/dx$) measurement with a resolution better than 6$\%$ for electrons from Bhabha scattering. The time-of-flight system (TOF) is composed of plastic scintillators with a time resolution of 80 (110) ps in the barrel (endcaps) and is used for charged particle identification. The CsI(Tl) electromagnetic calorimeter (EMC) measures 1~GeV energy photons with a resolution of 2.5$\%$ (5$\%$) in the barrel (endcaps) region.

Monte Carlo (MC) simulations are used to estimate the reconstruction efficiencies and study the backgrounds.
The signal MC samples
are generated using {\sc evtgen}~\cite{evtgen} using a $q^{2}$-dependent decay amplitude based on the assumption of a point-like meson, as described in Ref.~\cite{generator}, and an angular distribution based on that observed in data. An MC sample of generic $\psip$ decays, the so called ``inclusive MC sample", is used for the background studies. The production of the $\psip$ state is simulated by the {\sc kkmc}~\cite{kkmc} generator. The known decay modes of the $\psip$ are simulated by {\sc evtgen}~\cite{evtgen} according to the branching fractions reported in PDG~\cite{pdg}, while the unknown modes are simulated using the {\sc lundcharm}~\cite{lund} model.

Each charged track is required to have a point of closest approach to the interaction point (IP) that is less than 1~cm in the radial direction and less than 10 cm along the beam direction. The polar angle $\theta$ of the tracks must be within the fiducial volume of the MDC $(|\cos\theta|<0.93)$. Photons are reconstructed from isolated showers in the EMC which are at least $20^\circ$ away from the nearest charged track. The photon energy is required to be at least 25~MeV in the barrel region $(|\cos\theta|<0.8)$ or 50 MeV in the endcap region $(0.86<|\cos\theta|<0.92)$. In order to suppress electronic noise and energy depositions unrelated to the event, the time after the collision at which the photon is recorded in the EMC must be less than 700~ns.

Candidate events are required to have four charged tracks, with a sum of charges equal to zero, and at least one photon. The tracks with momentum larger than 1~GeV/$c$ are assumed to be leptons from $J/\psi$ decay. Otherwise they are considered as electrons from the $\psi'$ or $\chi_{cJ}$ decay. Leptons from the $J/\psi$ decay with EMC energy larger than 0.8 GeV are identified as electrons, otherwise as muons. The $\jpsi$ signal is identified by requiring the invariant mass of the lepton pair to be in the interval [3.08, 3.12] GeV/$c^{2}$. A vertex fit is performed on the four charged tracks to ensure the tracks originated from the IP. In order to reduce the background and improve the mass resolution, a four-constraint (4C) kinematic fit is performed by constraining the total four momentum to that of the initial beams. If there is more than one photon candidate in an event, all the photons are individually fit with the four leptons in the kinematic fit and only those with a fit $\chi^2< 40$ are retained. If two or more photons pass this criterion, only the one with the least $\chi^{2}$ is retained for further analysis.

A study of the $\psip$ inclusive MC sample shows that, after applying the above selection criteria, the main background comes from $\psip \too \gamma\chi_{cJ}, \chi_{cJ} \too \gamma J/\psi$ decays, where one photon converts into an $e^+e^-$ pair in the detector material. To suppress this background, a photon-conversion finder~\cite{conversion} is applied to reconstruct the photon-conversion vertex.
The distance from the point of the reconstructed conversion vertex to the $z$ axis, $R_{xy}$, is used to distinguish the photon conversion background from signal. By studying the MC samples $\psip \too \gamma\chi_{cJ}, \chi_{cJ} \too \gamma J/\psi$, the peaks around $R_{xy}=3$ cm and $R_{xy}=6$ cm match the positions of the beam pipe and the inner wall of the MDC~\cite{besiii}, respectively. We remove the events in $1.5$ cm$<R_{xy}<7.5$ cm to suppress the $\gamma$ conversion background. With this requirement, the $\gamma$ conversion background is negligible for the decays $\psip \too \EE\chi_{cJ}$ and is at the few percent  level for the decays $\chi_{cJ} \too \EE J/\psi$.

To remove the backgrounds from decays $\psip \too \eta/\pi^{0} J/\psi, \eta/\pi^{0} \too \gamma e^{+}e^{-}$, which have the same final state as signal events, a requirement 0.16 $< M(\gamma e^{+}e^{-}) <$ 0.50 GeV/$c^2$ is applied. By studying the data collected at $\sqrt{s}=3.65$ GeV, the contribution from the continuum process is found to be negligible.

\begin{figure}[htbp]
\begin{center}
\includegraphics[width=0.32\textwidth]{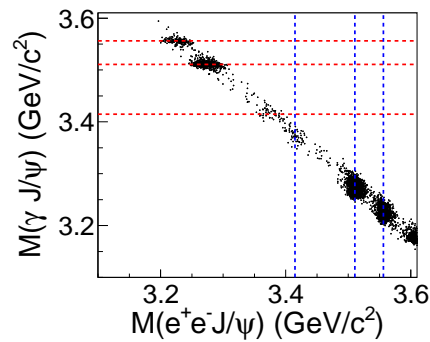}
\caption{(color online) Scatter plot of $M(\gamma J/\psi)$ versus $M(e^{+}e^{-}J/\psi)$ for data. The horizontal red dashed lines and vertical blue dashed lines indicate the positions of the $\chi_{cJ}$ masses in the $M(\gamma J/\psi)$ and $M(e^+e^- J/\psi)$ distributions, respectively.}
\label{fig:result}
\end{center}
\end{figure}

\begin{figure}[htbp]
\begin{center}
\includegraphics[width=0.23\textwidth]{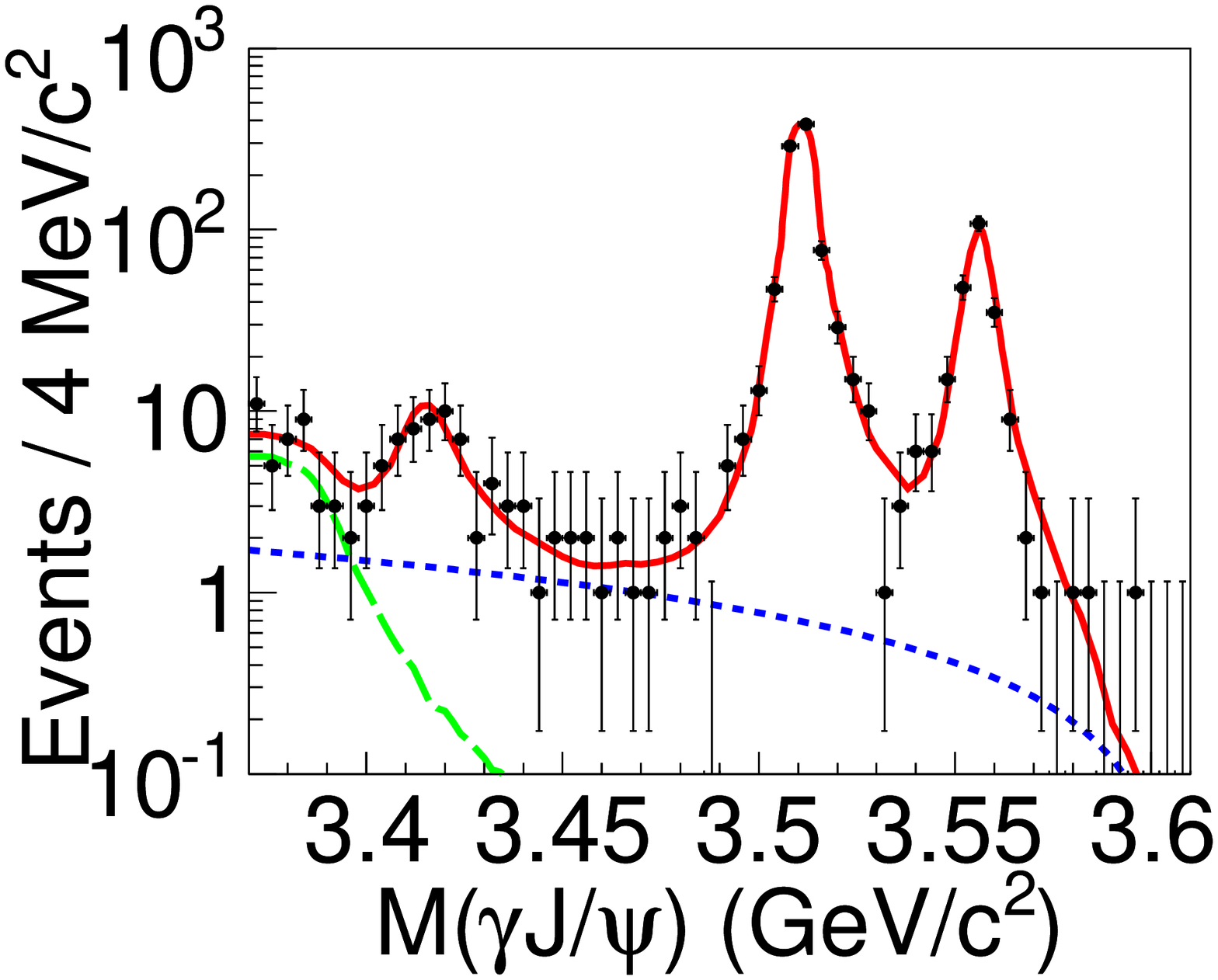}
\includegraphics[width=0.23\textwidth]{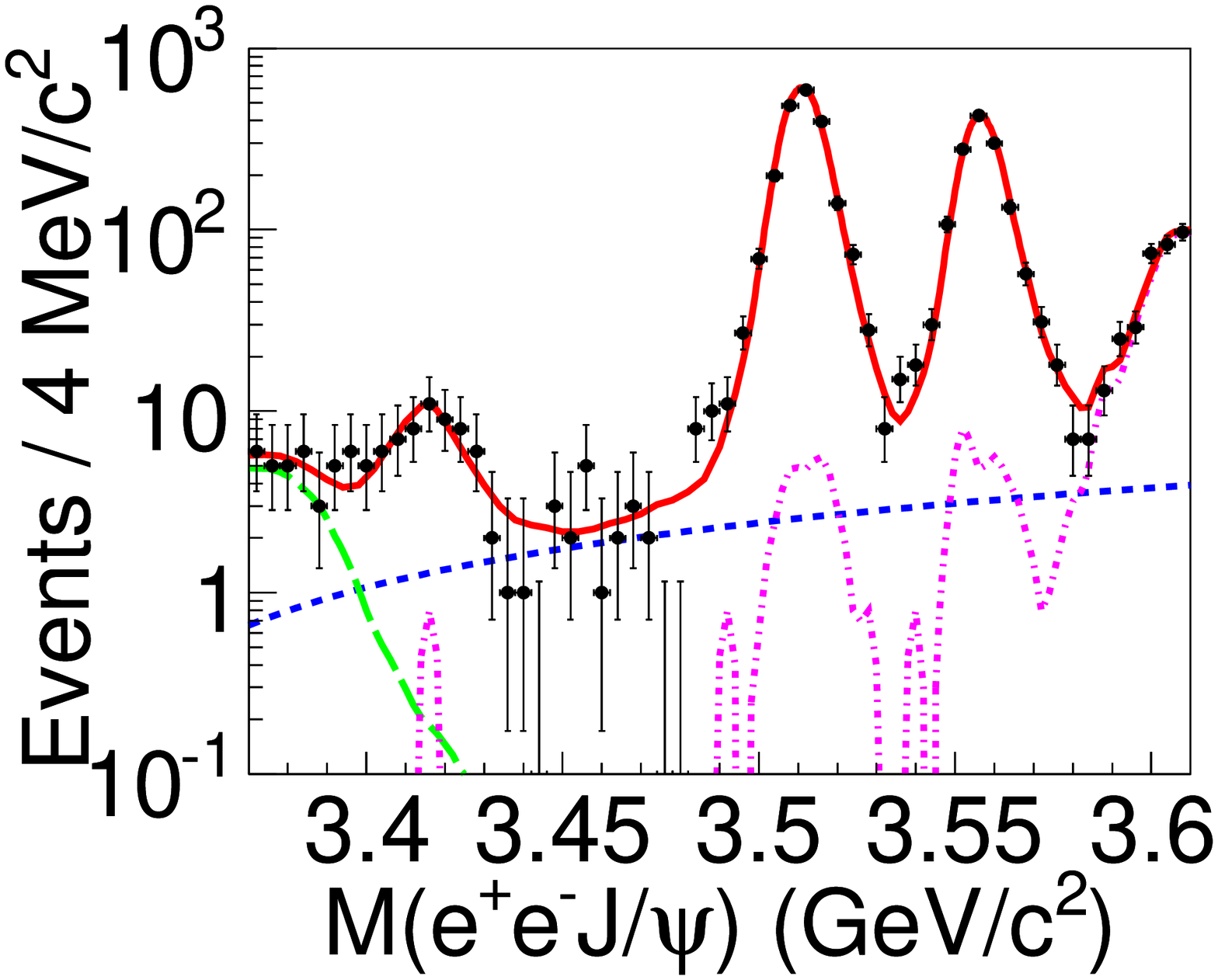}
\caption{ (color online) Data (points with error bars) distributions of (left) $M(\gamma J/\psi)$ and (right) $M(e^{+}e^{-}J/\psi)$. The red solid curve is the overall fit result, the green long-dashed curve is for the background (left) $\psip \too \gamma\chi_{c0}, \chi_{c0} \too e^{+}e^{-}J/\psi$ and (right) $\psip \too e^{+}e^{-}\chi_{c0}, \chi_{c0} \too \gamma J/\psi$, the blue dashed curve is for QED background, and the pink dashed-dotted curve in right plot is for the backgrounds from $\psip$ decays.}
\label{fig:fitresult}
\end{center}
\end{figure}

Figure~\ref{fig:result} shows the scatter plot of $M(\gamma J/\psi)$ versus $M(e^{+}e^{-} J/\psi)$ for the selected events from data; the corresponding one-dimensional projections are shown in Fig.~\ref{fig:fitresult}.
Clear $\chi_{cJ}$ signals are observed in the $M(\gamma J/\psi)$ and $M(e^{+}e^{-} J/\psi)$ distributions, corresponding to the decays $\psip \too e^{+}e^{-}\chi_{cJ}$ and $\chi_{cJ} \too e^{+}e^{-}J/\psi$, respectively.
The study of $\psip$ inclusive MC samples indicates that the dominant background is from the decay $\psip \too \pi^{+}\pi^{-}J/\psi, J/\psi \too (\gamma_{\rm FSR}) l^{+}l^{-}$, where $\gamma_{\rm FSR}$ is a photon due to final-state radiation; these events accumulate at $M(e^{+}e^{-}J/\psi)\sim 3.6~\mathrm{GeV}/c^2$.

Separate unbinned maximum likelihood fits are performed on the $M(\gamma J/\psi)$ and $M(e^{+}e^{-} J/\psi)$ distributions to extract the signal yields. We use the signal MC-determined shape, convoluted with a common Gaussian function, to describe the shapes of $\chi_{cJ}$ signals. The Gaussian function parametrizes any resolution difference between the data and MC simulation and its parameters are determined from the fit.

Two background components are considered in the fit to the $M(\gamma J/\psi)$ distribution. The first background is from the decay $\psip \too \gamma\chi_{c0}, \chi_{c0} \too e^{+}e^{-}J/\psi$, which corresponds to the peak at the lower edge of the  $M(\gamma J/\psi)$ region; it is described by a MC-determined shape with a fixed number of events based on the branching fraction obtained in this analysis. The second one is related to QED background ($\EE \too \ell^+\ell^-, \ell=e,\mu,\tau$) and is described by a first-order polynomial function in the fit.

In the fit to the $M(e^{+}e^{-}J/\psi)$ distribution, three  background components are considered. The first two are from the decay $\psip \too e^{+}e^{-}\chi_{c0}, \chi_{c0} \too \gamma J/\psi$, which corresponds to the enhancement at the lower edge of the $M(e^{+}e^{-} J/\psi)$ fit interval, and QED processes; the way these components are dealt with in this fit is analogous to the way they are dealt with in the $M(\gamma J/\psi)$ fit. The third background component is from inclusive $\psip$ decay, which includes the dominant one of $\psip \too \pi^{+}\pi^{-}J/\psi, J/\psi \too (\gamma_{\rm FSR}) l^{+}l^{-}$ decays and a small fraction from $\psip \too \gamma_{1} \chi_{cJ}, \chi_{cJ} \too \gamma_{2} J/\psi$, where $\gamma_{2}$ converts into an $e^{+}e^{-}$ pair.
In the fit, the shape of the third background component is assumed to be that reconstructed in the inclusive MC sample with the normalization determined from data.
The fit results are shown in Fig.~\ref{fig:fitresult} and the corresponding signal yields are summarized in Table~\ref{tab:branching}. For the six observed decay modes, the statistical significance of the yields are all larger than five standard deviations.

\begin{table*}[htbp]
\caption{Signal yields, detection efficiencies, the branching fractions and the ratios of the branching fractions. Here the first uncertainty is statistical and the second systematic. }
\label{tab:branching}
\begin{footnotesize}
\begin{tabular}{ c  c  c  c  c  c}
  \hline
  \hline
  Mode & Yields & Efficiency($\%$) & Branching fraction & $\frac{\mathcal{B}(\psip \too e^{+}e^{-}\chi_{cJ})}{\mathcal{B}(\psip \too \gamma\chi_{cJ})}$ & $\frac{\mathcal{B}(\chi_{cJ} \too e^{+}e^{-}J/\psi)}{\mathcal{B}(\chi_{cJ} \too \gamma J/\psi)}$ \\
  \hline
  $\psip \too e^{+}e^{-}\chi_{c0}$ & $48\pm10$  & 6.06 & $(11.7 \pm 2.5 \pm 1.0)\times10^{-4}$ & $(9.4 \pm 1.9 \pm 0.6)\times10^{-3}$ & $-$ \\
  $\psip \too e^{+}e^{-}\chi_{c1}$ & $873\pm30$ & 5.61 & $(8.6 \pm 0.3 \pm 0.6)\times10^{-4}$  & $(8.3 \pm 0.3 \pm 0.4)\times10^{-3}$ & $-$ \\
  $\psip \too e^{+}e^{-}\chi_{c2}$ & $227\pm16$ & 3.19 & $(6.9 \pm 0.5 \pm 0.6)\times10^{-4}$  & $(6.6 \pm 0.5 \pm 0.4)\times10^{-3}$ & $-$ \\

  $\chi_{c0} \too e^{+}e^{-}J/\psi$ & $56\pm11$   & 6.95 & $(1.51 \pm 0.30 \pm 0.13)\times10^{-4}$ & $-$ & $(9.5 \pm 1.9 \pm 0.7)\times10^{-3}$  \\
  $\chi_{c1} \too e^{+}e^{-}J/\psi$ & $1969\pm46$ & 10.35 & $(3.73 \pm 0.09 \pm 0.25)\times10^{-3}$ & $-$ & $(10.1 \pm 0.3 \pm 0.5)\times10^{-3}$ \\
  $\chi_{c2} \too e^{+}e^{-}J/\psi$ & $1354\pm39$ & 11.23 & $(2.48 \pm 0.08 \pm 0.16)\times10^{-3}$ & $-$ & $(11.3 \pm 0.4 \pm 0.5)\times10^{-3}$ \\
  \hline
  \hline
\end{tabular}
\end{footnotesize}
\end{table*}

%The $\chi^{2}/$ndf are $26.4/22$ and $51.1/36$ for the fit projections on the binned $M(\gamma J/\psi)$ and $M(e^{+}e^{-}J/\psi)$ distributions, respectively. Here, ndf is the number of degrees of freedom, which is calculated as the number of bins considered less the number of free parameters in the fit. Only bins with more than seven entries are considered, otherwise, the events in a bin are migrated to a neighboring bin.

The branching fractions $\mathcal{B}(\psip \too e^{+}e^{-}\chi_{cJ})$ and $\mathcal{B}(\chi_{cJ} \too e^{+}e^{-}J/\psi)$ are calculated according to
\begin{equation}
    \mathcal{B}=\frac{N_{\mathrm{sig}}}{N_{\psip}\cdot{\epsilon}\cdot{\mathcal{B}_{\mathrm{radiative}}}\cdot{\mathcal{B}(J/\psi \too l^{+}l^{-})}},
\end{equation}
where $N_{\mathrm{sig}}$ is the corresponding number of signal events extracted from the fit, $N_{\psip}$ is the total number of $\psip$ events, $\epsilon$ is the selection efficiency determined from the signal MC samples, $\mathcal{B}_{\mathrm{radiative}}$ is the branching fraction of the radiative transitions $\psip \too \gamma\chi_{cJ}$ or $\chi_{cJ} \too \gamma J/\psi$, and $\mathcal{B}(J/\psi \too l^{+}l^{-})$ is the decay branching fraction of $J/\psi \too l^{+}l^{-}$. All the branching fractions used are taken from Ref.~\cite{pdg}. The resultant branching fractions of $\psip \too e^{+}e^{-}\chi_{cJ}$ and $\chi_{cJ} \too e^{+}e^{-}J/\psi$ are listed in Table~\ref{tab:branching}.

\begin{figure}[htbp]
\begin{center}
\begin{overpic}[width=0.23\textwidth]{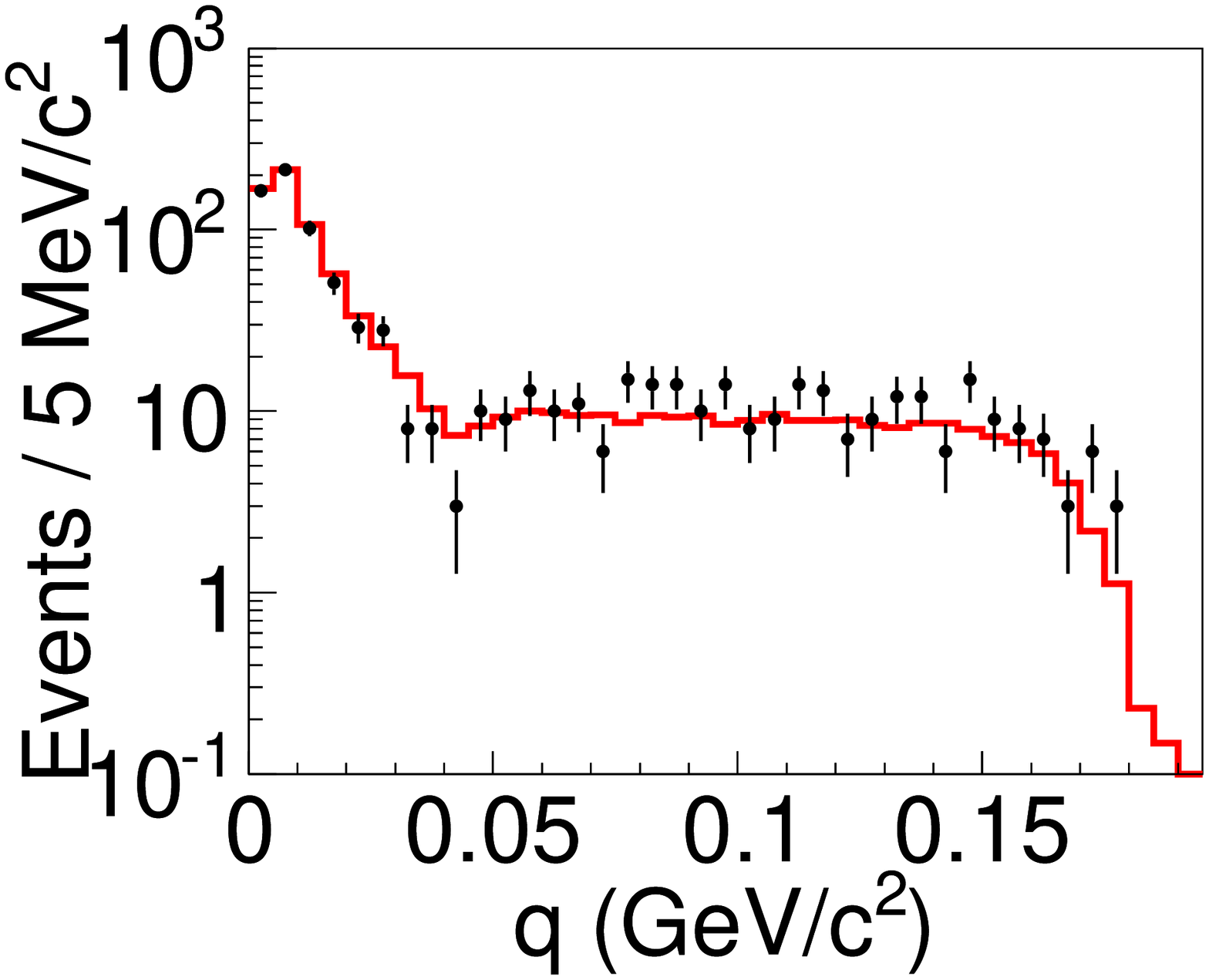}
\put(97,75){(a)}
\end{overpic}
\begin{overpic}[width=0.23\textwidth]{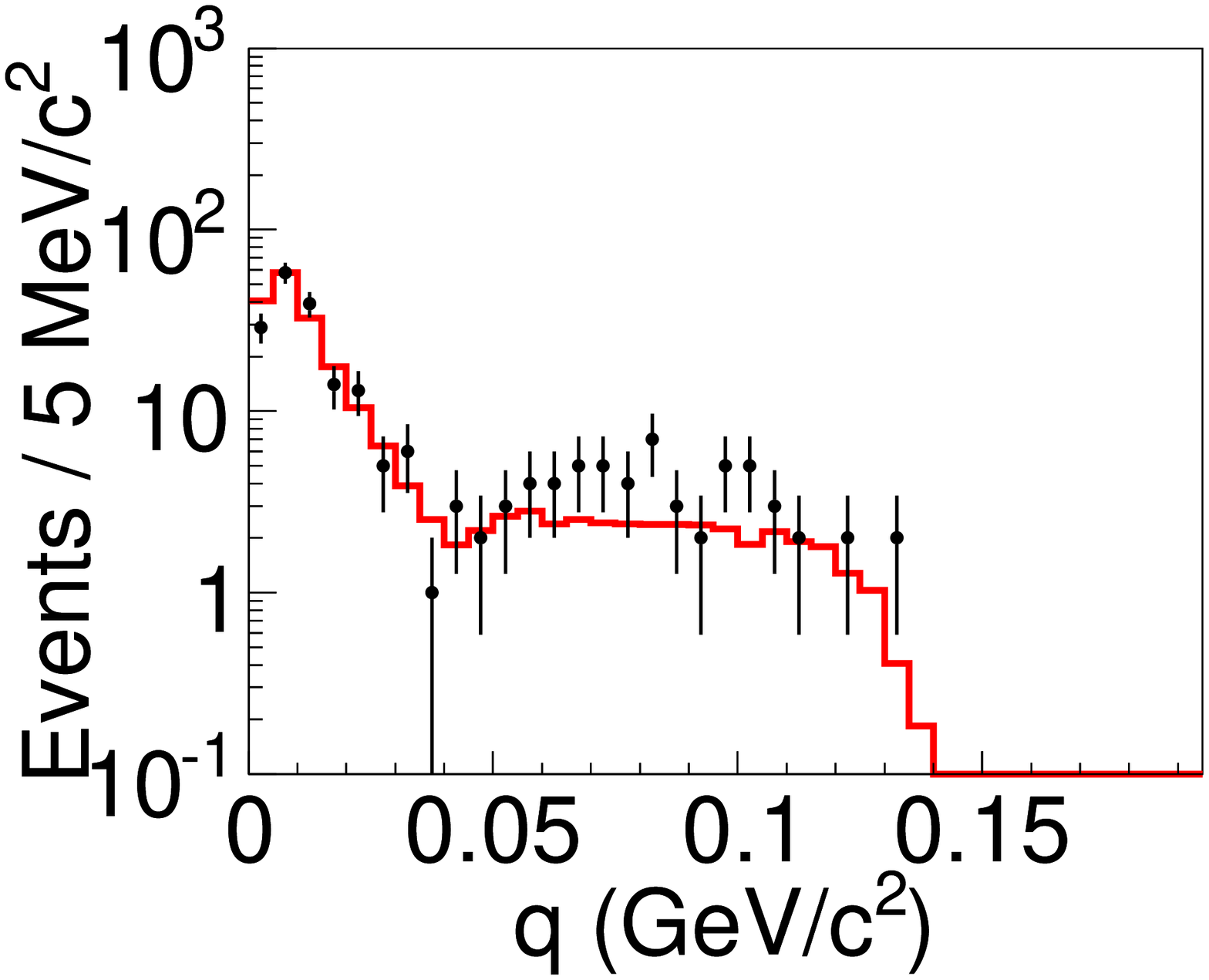}
\put(97,75){(b)}
\end{overpic}
\begin{overpic}[width=0.23\textwidth]{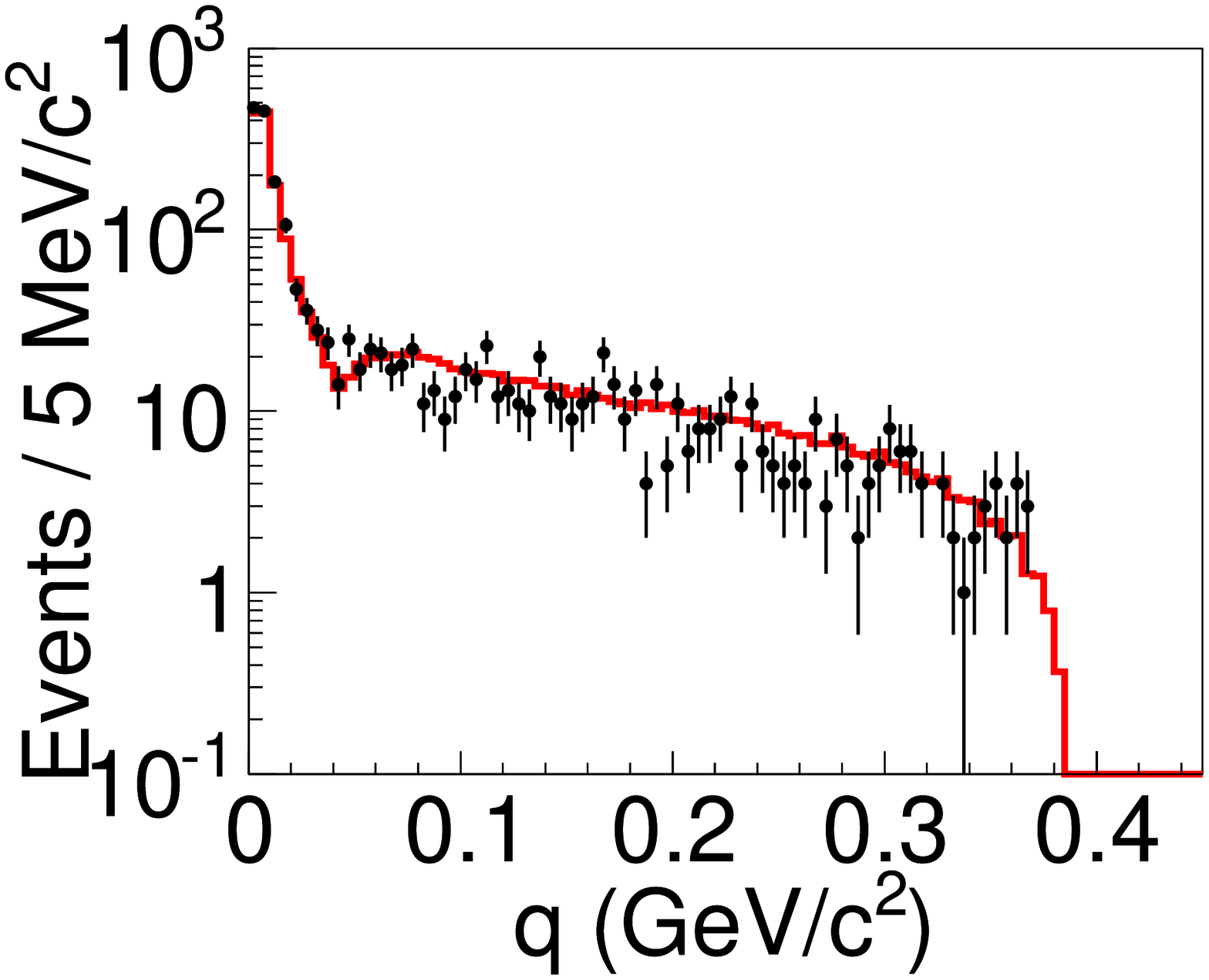}
\put(97,75){(c)}
\end{overpic}
\begin{overpic}[width=0.23\textwidth]{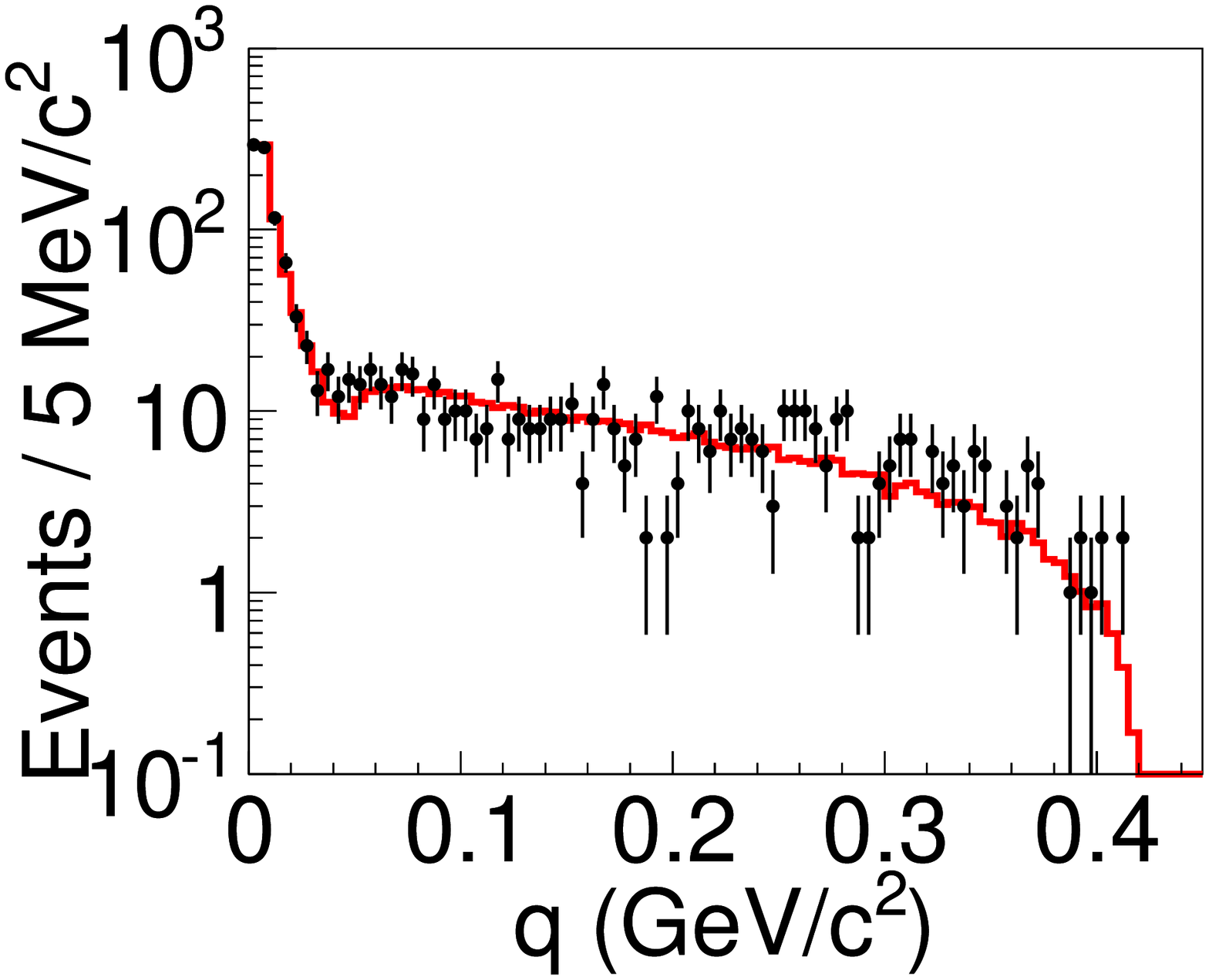}
\put(97,75){(d)}
\end{overpic}
\caption{Data to MC simulation comparisons of $q$ distribution for the decays (a) $\psip \too e^{+}e^{-}\chi_{c1}$, (b) $\psip \too e^{+}e^{-}\chi_{c2}$, (c) $\chi_{c1} \too e^{+}e^{-}J/\psi$ and (d) $\chi_{c2} \too e^{+}e^{-}J/\psi$. The points with error bars are data and the red histograms are for the signal MC simulation.}
\label{fig:eemass}
\end{center}
\end{figure}

Figure~\ref{fig:eemass} shows comparisons of the $q$ distributions in data and MC simulation for the decays $\psip \too e^{+}e^{-}\chi_{c1,2}$ and $\chi_{c1,2} \too e^{+}e^{-}J/\psi$, where the $\chi_{c1}$ and $\chi_{c2}$ signals are extracted requiring a mass within [3.49,3.53] and [3.54,3.58] GeV/$c^2$, respectively; with these criteria the backgrounds are expected to be less than 2$\%$. The data are in reasonable agreement with the MC simulation generated using the model described in Ref.~\cite{generator}.

The systematic uncertainties for the branching fraction measurement arise from the following sources: track reconstruction, photon detection, kinematic fitting, $J/\psi$ mass criteria, $M(\gamma e^{+}e^{-})$ requirement, $\gamma$ conversion vetoing, fit procedure,  angular distributions, the total number of $\psip$ events and the branching fractions of the cascade decays. All uncertainties are discussed in detail below.

The difference in the tracking efficiency between data and the MC simulation, for each charged track, is estimated to be $1.0\%$~\cite{trackeff}, which results in a $4.0\%$ systematic uncertainty for all modes. The uncertainty on the photon-detection efficiency is derived from a control sample of $J/\psi \too \rho^{0}\pi^{0}$  decays and is $1.0\%$ per photon~\cite{rhopi}.

In the 4C kinematic fit, the helix parameters of charged tracks are corrected to reduce the discrepancy between data and the MC simulation as described in Ref.~\cite{helix}. The correction factors are obtained by studying a control sample of $\psip \too \pi^{+}\pi^{-}J/\psi, J/\psi \too l^{+}l^{-}$ decays. To determine the systematic uncertainty from this source, we determine the efficiencies from the MC samples without the helix correction; the resulting differences with respect to the nominal values are taken as the systematic uncertainties.

The uncertainty associated with the $J/\psi$ mass requirement is 1.0\%, which is determined by studying a control sample of $\psip \too \eta J/\psi, \eta \too \gamma\gamma$ (where one $\gamma$ undergoes conversion to an $e^+e^-$ pair) or $\eta \too \gamma e^{+}e^{-}$ decays. The systematic uncertainty related to the $M(\gamma e^{+}e^{-})$ interval used is studied by varying the edges of the interval by $\pm 5$ MeV$/c^{2}$. The largest difference with the nominal value is taken as the systematic uncertainty from this source.

To study the systematic uncertainty related to the $\gamma$ conversion background veto, we compare the efficiencies of $\gamma$ conversion veto between data and the MC simulation in control samples of $\psip \too \gamma\chi_{c1,2}, \chi_{c1,2} \too e^{+}e^{-} J/\psi$ decays. The efficiency of the $\gamma$ conversion veto
is the ratio of the signal yields determined by fitting the $M(e^{+}e^{-})$ distribution with and without the $\gamma$ conversion veto applied. A relative difference between data and simulation of $1.4\%$ is found and assigned as the systematic uncertainty.

The sources of uncertainty in the fit procedure include the fit range and the signal and background parametrization. The uncertainty related with the fit range is obtained by varying the limits of the fit range by $\pm$5 MeV/$c^{2}$. The largest difference in the signal yields with respect to the nominal values is taken as the systematic uncertainty. In the nominal fit, the signal shapes are described with the signal MC simulated shapes convoluted with a Gaussian function. An alternative fit is performed by fixing the signal shapes to those of MC simulation. The resultant change in the signal yields is taken as the systematic uncertainty.
The uncertainty associated with the background shape is estimated by an alternative fit replacing the first order polynomial function with a second order polynomial function for the background shape,
the resultant change in the signal yields is taken as the systematic uncertainty.

\begin{figure}[htbp]
\begin{center}
\begin{overpic}[width=0.23\textwidth]{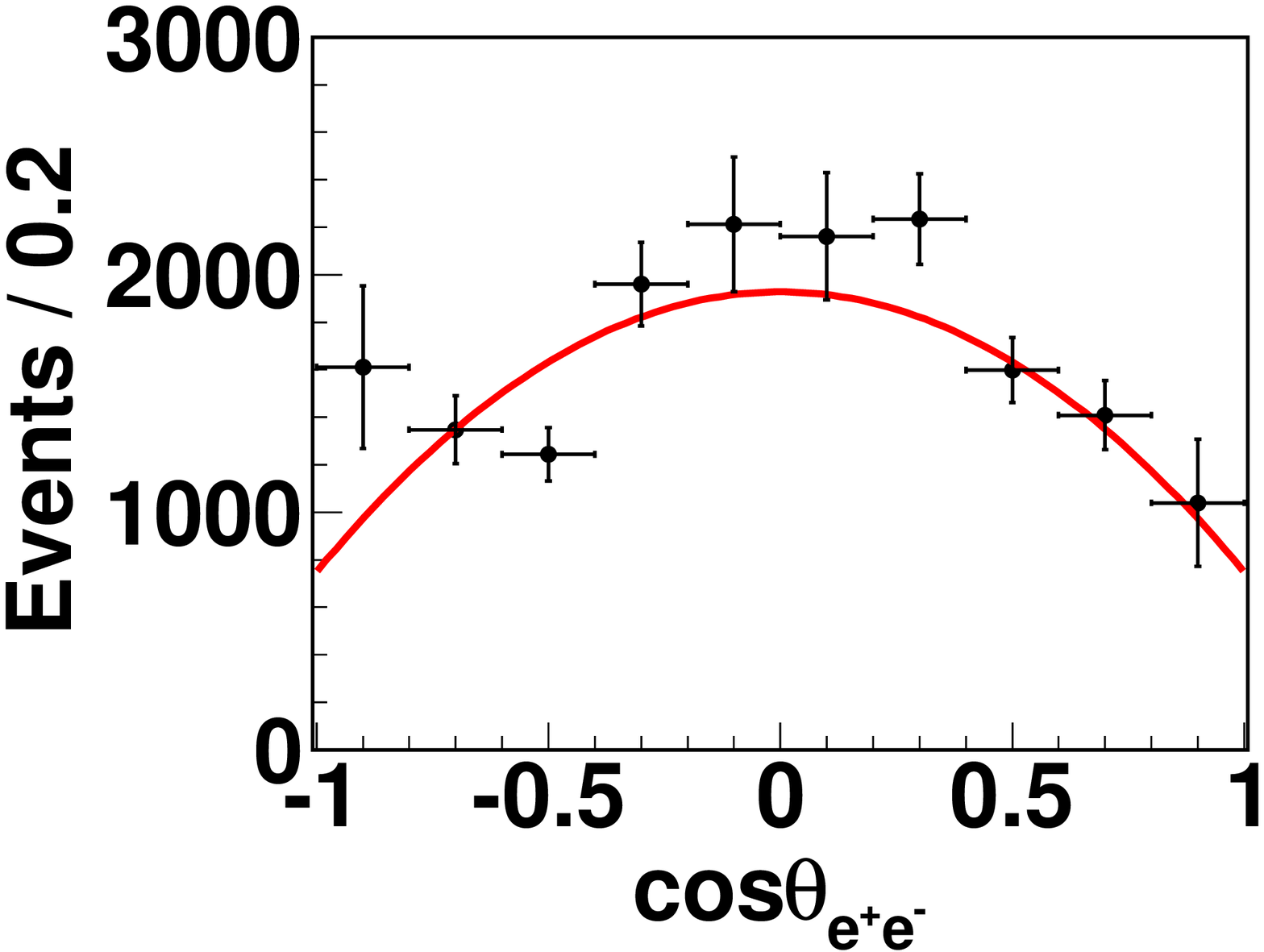}
\put(97,70){(a)}
\end{overpic}
\begin{overpic}[width=0.23\textwidth]{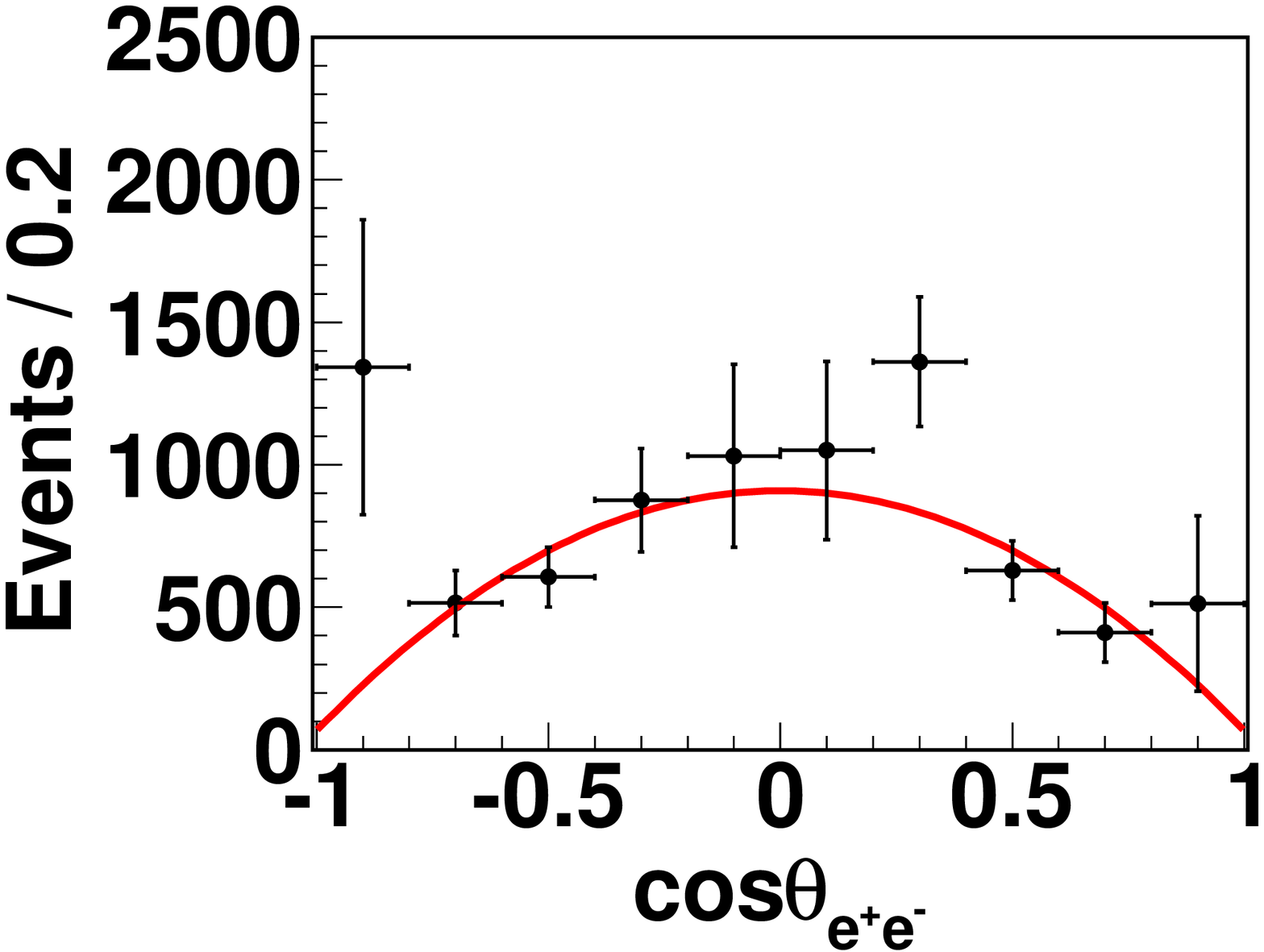}
\put(97,70){(b)}
\end{overpic}
\begin{overpic}[width=0.23\textwidth]{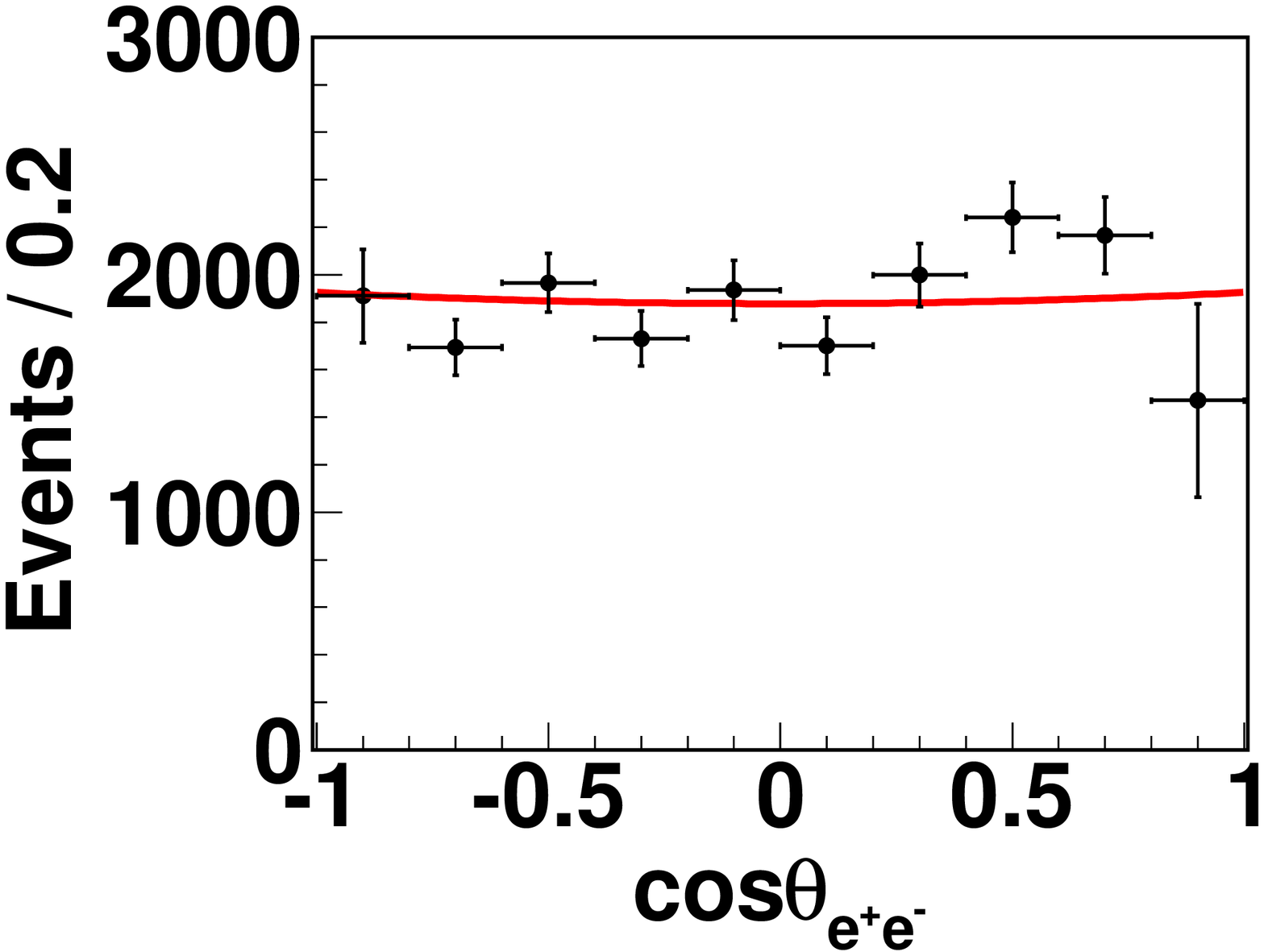}
\put(97,70){(c)}
\end{overpic}
\begin{overpic}[width=0.23\textwidth]{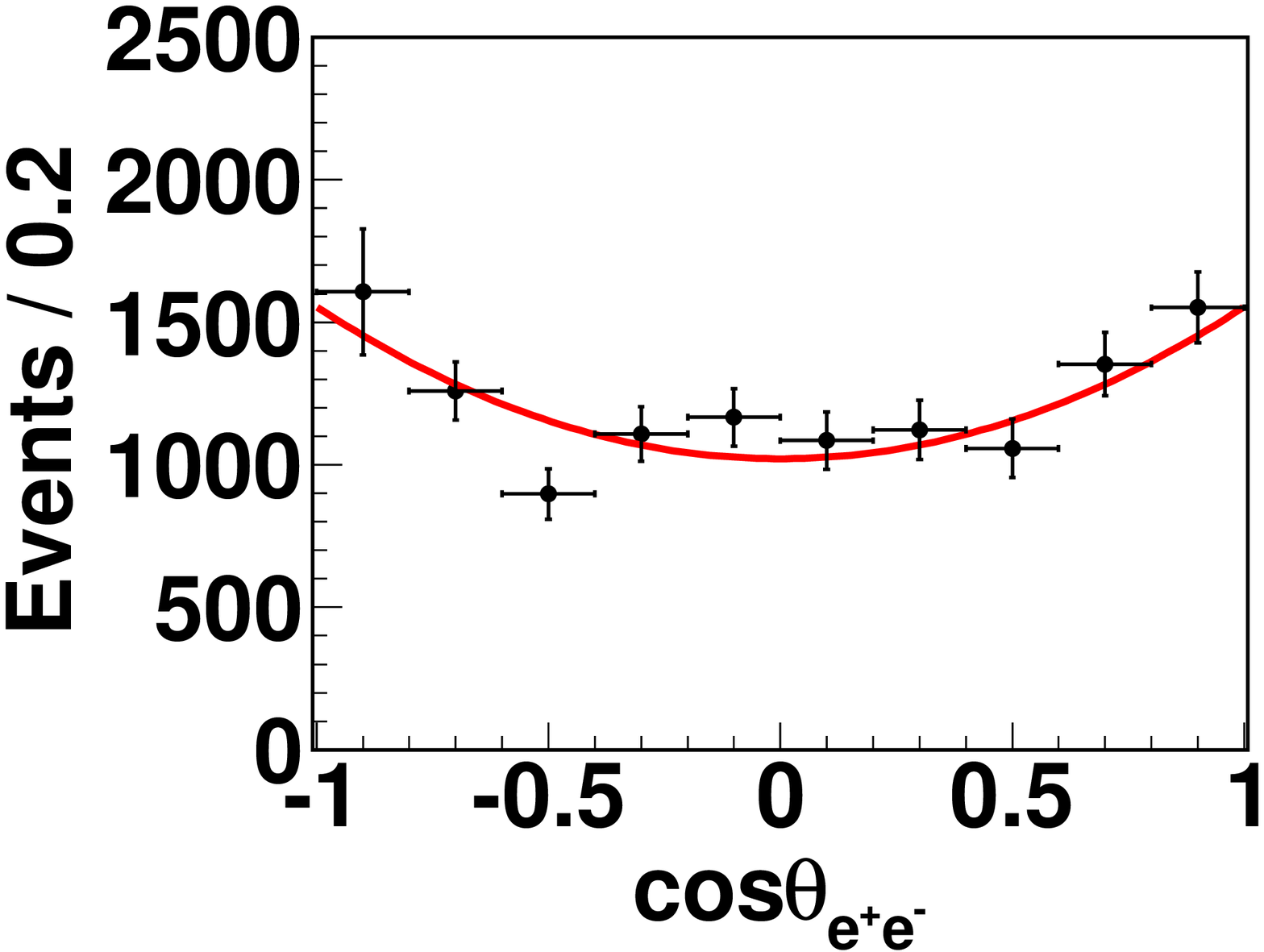}
\put(97,70){(d)}
\end{overpic}
\caption{Distributions of efficiency corrected  cos$\theta_{e^{+}e^{-}}$ for the decays (a) $\psip \too e^{+}e^{-}\chi_{c1}$, (b) $\psip \too e^{+}e^{-}\chi_{c2}$, (c) $\chi_{c1} \too e^{+}e^{-}J/\psi$ and (d) $\chi_{c2} \too e^{+}e^{-}J/\psi$. The red line is the fit to $1+\alpha\cos^{2}\theta_{e^{+}e^{-}}$.}
\label{fig:angle}
\end{center}
\end{figure}

The distribution of $e^+e^-$ pair's helicity angle in its mother rest frame $\theta_{e^+e^-}$ may affect the detector efficiency, where $\theta_{e^+e^-}$ is the polar angle of $e^+e^-$ pair in the colliding beams rest frame with the $z$ axis pointing in the positron beam direction.
The efficiency corrected $\cos\theta_{e^{+}e^{-}}$ distributions are shown in Fig.~\ref{fig:angle} for the decays $\psip \too e^{+}e^{-}\chi_{c1,2}$ and $\chi_{c1,2} \too e^{+}e^{-}J/\psi$; each distribution is fit with a $1+\alpha\cos^{2}\theta_{e^{+}e^{-}}$ function.
The resultant $\alpha$ values are $-0.6\pm0.2$, $-0.9\pm0.3$, $0.0\pm0.2$ and $0.5\pm0.2$ for the decays $\psip \too e^{+}e^{-}\chi_{c1}$, $\psip \too e^{+}e^{-}\chi_{c2}$, $\chi_{c1} \too e^{+}e^{-}J/\psi$ and $\chi_{c2} \too e^{+}e^{-}J/\psi$, respectively.
The measured $\alpha$ central values are incorporated in the nominal MC simulations. To take into account any effect on the detection efficiencies due to an incorrect simulation of the $\cos\theta_{e^+e^-}$ distribution, alternative MC samples are generated with $\alpha$ varied by $\pm 1$ standard deviation and the efficiencies are determined. The differences with the nominal efficiencies are taken as the systematic uncertainties from this source.
In the decays $\psip \too e^{+}e^{-}\chi_{c0}$ and $\chi_{c0} \too e^{+}e^{-}J/\psi$, the $\cos\theta_{e^+e^-}$ distribution is not extracted directly from the data due to the limited statistics. The theoretical expectations for $\alpha$ are 1 and 0 for $\psip \too e^{+}e^{-}\chi_{c0}$ and $\chi_{c0} \too e^{+}e^{-}J/\psi$, respectively, which are used to generate the nominal MC simulation. The systematic uncertainty is estimated using the difference in efficiency when alternative MC samples with $\alpha=0$ for $\psip \too e^{+}e^{-}\chi_{c0}$ and $\alpha=1$ for $\chi_{c0} \too e^{+}e^{-}J/\psi$ are used.

The total number of $\psip$ events is measured to within 0.7\% by using the inclusive hadronic events~\cite{totalnumber, totalnumber2}.
The uncertainties of the branching fractions in the cascade decays are taken from Ref.~\cite{pdg}.

The effect of other potential systematic uncertainty sources are considered, such as uncertainties on the generated $q$ distributions, the trigger efficiency, and the simulation of the event time, but are all found to be negligible.
Table~\ref{tab:sumerror} summarizes all individual systematic uncertainties, and the overall uncertainties are the quadrature sums of the individual ones, assuming they are independent.

\begin{table}[htbp]
\caption{Summary of systematic uncertainties (in $\%$). }
\label{tab:sumerror}
\begin{footnotesize}
\begin{tabular}{ cccccccc  }
  \hline
  \hline
   &\multicolumn{3}{c}{~$\psip \too e^{+}e^{-}\chi_{cJ}$~~} & \multicolumn{3}{c}{$\chi_{cJ} \too e^{+}e^{-}J/\psi$~~} \\
   &~~~$\chi_{c0}$~~ &~~$\chi_{c1}$~~&~~$\chi_{c2}$~~&~~$\chi_{c0}$~~&~~~$\chi_{c1}$~~&~~$\chi_{c2}$~~ \\
  \hline
  Tracking                    & 4.0& 4.0& 4.0 & 4.0& 4.0& 4.0 \\

  Photon                      & 1.0& 1.0&  1.0 & 1.0& 1.0& 1.0 \\

  Kinematic fit               & 1.6& 1.4& 1.4 & 1.8& 2.2& 2.4 \\

  $J/\psi$ mass window        & 1.0& 1.0& 1.0 & 1.0& 1.0& 1.0 \\

  $M(\gamma e^{+}e^{-})$      & 2.7& 1.2& 1.0 & 0.7& 2.2& 0.4 \\

  $\gamma$ conversion vetoing & 1.4& 1.4& 1.4 & 1.4& 1.4& 1.4 \\

  Fit  Range                  & 2.2& 0.2& 0.3 & 4.7& 0.1& 0.2 \\

  Signal shape                & 0.4& 0.1& 0.1 & 2.2& 0.2& 0.5 \\

  Background shape            & 2.2& 0.2& 0.3 & 0.1& 0.1& 0.2 \\

  Angular distribution        & 3.9& 2.1& 3.3 & 3.6& 1.6& 1.0 \\

  Number of $\psip$           & 0.7& 0.7& 0.7 & 0.7& 0.7& 0.7 \\

  Branching fractions         & 4.8& 3.6& 5.5 & 2.8& 3.3& 3.5 \\

  \hline

  sum                         & 8.9& 6.5& 8.1 & 8.5& 6.6& 6.3 \\
  \hline
  \hline
\end{tabular}
\end{footnotesize}
\end{table}

In summary, using a data sample of $4.479 \times 10^{8}$ $\psip$ events collected with the BESIII detector operating at the BEPCII collider,
the decays $\psip \too e^{+}e^{-}\chi_{cJ}$ and $\chi_{cJ} \too e^{+}e^{-}J/\psi$ are observed for the first time, and the corresponding branching fractions
are measured and the values are given in Table~\ref{tab:branching}.
The ratios of branching fractions $\frac{\mathcal{B}(\psip \too e^{+}e^{-}\chi_{cJ})}{\mathcal{B}(\psip \too \gamma\chi_{cJ})}$ and $\frac{\mathcal{B}(\chi_{cJ} \too e^{+}e^{-}J/\psi)}{\mathcal{B}(\chi_{cJ} \too \gamma J/\psi)}$ are also obtained by incorporating the BESIII results of the product of branching fractions $\mathcal{B}(\psip \too \gamma\chi_{cJ})\cdot\mathcal{B}(\chi_{cJ} \too \gamma J/\psi)$ in Ref.~\cite{gaoq}, as listed in Table~\ref{tab:branching}.
The common systematic uncertainties related to efficiency and branching fractions cancel in the calculation.
The measured $q^{2}$ distributions are consistent with those of the signal MC simulation based on the assumption of a point-like meson \cite{generator}. This first observation of the $q^2$-dependent charmonium EM Dalitz transitions can help understand the discrepancy between the experimental measurements~\cite{pdg} and the theoretical predictions~\cite{model1,model2,model3,model4} of the $\psip \too \gamma\chi_{cJ}$ branching fractions. The experimental methods applied here for the first study of charmonium Dalitz decays are likely to be of use for similar studies of the $X(3872)$. It is hoped that this experimental work will spur new theoretical development on use of charmonium Dalitz decays to address questions such as the nature of exotic charmonium.

%%%%%%%%%%%%%%%%%%%%%%%%%%%%%%%%%%%%%%
%%%%%%%%%%%%%%%%%%%%%%%%%%%%%%%%%%%%%%

The BESIII collaboration thanks the staff of BEPCII and the IHEP
computing center for their strong support. This work is supported in
part by National Key Basic Research Program of China under Contract
No. 2015CB856700; National Natural Science Foundation of China (NSFC)
under Contracts Nos. 11125525, 11235011, 11322544, 11335008, 11425524,
11521505, 11575198; the Chinese Academy of Sciences (CAS) Large-Scale
Scientific Facility Program; the CAS Center for Excellence in Particle
Physics (CCEPP); the Collaborative Innovation Center for Particles and
Interactions (CICPI); Joint Large-Scale Scientific Facility Funds of
the NSFC and CAS under Contracts Nos. 11179007, U1232201, U1332201;
CAS under Contracts Nos. KJCX2-YW-N29, KJCX2-YW-N45; 100 Talents
Program of CAS; National 1000 Talents Program of China; INPAC and
Shanghai Key Laboratory for Particle Physics and Cosmology; German
Research Foundation DFG under Contracts Nos. Collaborative Research
Center CRC-1044, FOR 2359; Istituto Nazionale di Fisica Nucleare, Italy;
Koninklijke Nederlandse Akademie van Wetenschappen (KNAW) under
Contract No. 530-4CDP03; Ministry of Development of Turkey under
Contract No. DPT2006K-120470; Russian Foundation for Basic Research
under Contract No. 14-07-91152; The Swedish Resarch Council;
U.S. Department of Energy under Contracts Nos. DE-FG02-05ER41374,
DE-SC-0010504, DE-SC0012069, DESC0010118; U.S. National Science
Foundation; University of Groningen (RuG) and the Helmholtzzentrum
fuer Schwerionenforschung GmbH (GSI), Darmstadt; WCU Program of
National Research Foundation of Korea under Contract
No. R32-2008-000-10155-0.

\end{document}